\documentclass[12pt]{article}
\usepackage{graphicx}
\usepackage[bookmarksnumbered,colorlinks,plainpages]{hyperref}
\usepackage{amssymb,amsfonts,amsmath}
\usepackage{makeidx}
\usepackage{amssymb,amsfonts,amsmath}
\usepackage{verbatim}

\begin{document}

\title{Is it possible to formulate least action principle for dissipative systems?}
\author{Qiuping A. Wang and Ru Wang\\
{\small $^1$IMMM, UMR CNRS 6283, Universit\'e du Maine, 72085 Le Mans, France}}

\date{}

\maketitle

\date{}

\begin{abstract}
A longstanding open question in classical mechanics is to formulate the least action principle for dissipative systems. In this work, we give a general formulation of this principle by considering a whole conservative system including the damped moving body and its environment receiving the dissipated energy. This composite system has the conservative Hamiltonian $H=K_1+V_1+H_2$ where $K_1$ is the kinetic energy of the moving body, $V_1$ its potential energy and $H_2$ the energy of the environment. The Lagrangian can be derived by using the usual Legendre transformation $L=2K_1+2K_2-H$ where $K_2$ is the total kinetic energy of the environment. An equivalent expression of this Lagrangian is $L=K_1-V_1-E_d$ where $E_d$ is the energy dissipated by the friction from the moving body into the environment from the beginning of the motion. The usual variation calculus of least action leads to the correct equation of the damped motion. We also show that this general formulation is a natural consequence of the virtual work principle.
\end{abstract}


\vspace{2 cm}

\section{Introduction}
The Least Action Principle (LAP) is one of the most valuable heritages from the classical mechanics\cite{Maupertuis,Arnold,Stoltzner}. The fact that the formulation of the whole classical physics as well as of the quantum theory in its path integral formalism\cite{Feynman} could be based on or related to this single mathematical rule gives to LAP a fundamental priority to all other visibly different principles, empirical laws and differential equations in different branches of physics. This priority of LAP has nourished two major hopes or ambitions of physicists. The first one is the (rather controversial) effort to deepen the understanding of nature through this principle and to search for the fundamental meaning of its exceptional universality in physics\cite{Stoltzner,Mach,Lanczos}. The second one is to extend it to more domains such as thermodynamics and statistical mechanics (with the pioneer efforts of Boltzmann, Helmholtz and Hertz\cite{Stengers}), stochastic dynamics (e.g., large deviation theory\cite{Freidlin} and stochastic mechanics\cite{Yasue}), and dissipative mechanical systems\cite{Sieniutycz,Vujanovic,Herrera}. The present work is following this last effort to formulate LAP for dissipative mechanical systems.

LAP was originally formulated only for Hamiltonian system\cite{Arnold}, i.e., the sum $H=K+V$ of kinetic energy $K$ and potential energy $V$ of the considered system satisfies the Hamiltonian equations. For Hamiltonian systems, any real trajectory between two given configuration points must satisfy the LAP, a vanishing first variation of the action $A$ created by tiny deformation of the trajectory\cite{Arnold,Lanczos}:
\begin{equation}    \label{e1}
\delta A=\delta \int_0^{t_b} Ldt=\int_0^{t_b} \delta Ldt=0
\end{equation}
where the action $A=\int_0^{t_b} Ldt$ is a time integral of the Lagrangian $L=K-V$ on the trajectory over a fixed time period $t_b$. One of the important results of this variational calculus is the Euler-Lagrange equation given by\cite{Lanczos} (for one freedom $x$)
\begin{equation}    \label{e2}
\frac{d}{dt}\left(\frac{\partial L}{\partial \dot{x}}\right)-\frac{\partial L}{\partial x}=0
\end{equation}
where $\dot{x}$ is the velocity. In many cases when $H$ and $L$ do not depend on time explicitly, a Hamiltonian system is energy conservative. A problem of variation takes place for nonconservative systems having, say, a dissipative force $f_d$ which is introduced into the equation of motion in this way $\frac{d}{dt}\left(\frac{\partial L}{\partial \dot{x}}\right)-\frac{\partial L}{\partial x}=f_d$. This is equivalent to write $\int_0^{t_b}(\delta L +f_d\delta x)dt=0$. However, it is impossible to define an action integral with a single (Lagrangian) function satisfying Eq.(\ref{e2}). Hence LAP is absent for dissipative systems.

There has been a longstanding effort in classical mechanics to formulate LAP, or variational calculus in general, for nonconservative or dissipative system\cite{Vujanovic}. This formulation is essential for this fundamental principle of physics to be applied to a wide range of systems in nature that are not conservative, to be well understood and to be related to many scientific principles relative to the energy dissipation\cite{Vujanovic,Goldstine,Sieniutycz}. As far as we know, the first variational calculus applied to damped motion dates back to Euler's work in 1744 for the brachistochrone (shortest time) problem with friction\cite{Goldstine}. More recently, Rayleigh\cite{Goldstein} has introduced a `dissipative function' $D=\frac{1}{2}\zeta \dot{x}^2$, for the special case of the Stokes' law  $f_d=-m\zeta\dot{x}$, to write $\frac{d}{dt}\left(\frac{\partial L}{\partial \dot{x}}\right)+\frac{\partial D}{\partial \dot{x}}-\frac{\partial L}{\partial x}=0$, where $\zeta$ is the drag constant and $m$ the mass of the damped body. Nevertheless, LAP is not recovered since there is no Lagrangian for defining an action which satisfies Eq.(\ref{e1}). Other major propositions include the Bateman approach\cite{Bateman} to introduce complementary variables and equations, the definition of dissipative Lagrangian by multiplying the non dissipative one with an exponential factor $exp(\zeta t)$\cite{Sanjuan} where $t$ is the time, the fractional derivative formulation\cite{Riewe}, and the pseudo-Hamiltonian mechanics\cite{Duffin} where a parameter was introduced to characterize the degree of dissipation. The reader is referred to the reviews in \cite{Sieniutycz,Vujanovic,Gray,Riewe,Duffin} about the details of these propositions. In general, the Lagrangian in these solutions is not unique and has no energy connection like $L=K-V$ (see for instance the quasi-Lagrangian $L=e^{\zeta t}(K-V)$ and the corresponding quasi-Hamiltonian $H=e^{-\zeta t}K+e^{\zeta t}V$ for damped harmonic oscillator\cite{Sanjuan}). Hence no variational or optimal calculus was possible in general form\cite{Sieniutycz,Vujanovic,Gray}.

A common character of these previous works is that the damped body is the only object taken into account in the calculations as if it was isolated. However, a dissipative system is always coupled to an environment and loses energy into the latter, an integral part of the motion. As far as this lost energy is not considered, the quasi-Lagrangian function of the damped body inevitably loses energy connection and generic optimal characters\cite{Sieniutycz,Vujanovic} as mentioned above.

The aim of this work is to formulate LAP in a general way for classical system, dissipative or not, with an energy connected and unique Lagrangian, and its relation with a conservative Hamiltonian (Legendre transform). It will be shown that the three conventional formulations of analytical mechanics, i.e., the Hamiltonian, Lagrangian and the Hamilton-Jacobi equations, are all preserved.

\section{The conservative Hamiltonian}

Our basic idea is to consider the damped moving body and its environment, coupled to each other by dissipative force, as a whole conservative system. The total Hamiltonian includes the instantaneous kinetic and the potential energy of the body, as well as the mechanical energy that is transformed into heat or other forms of energy (noises, vibration, electromagnetic radiation etc.) in the environment. The body (system 1) is large with respect to the particles of the environment and moves along the axis $x$ with velocity $\dot{x}$. Its environment (system 2 composed of $N$ particles with position $x_i$ and velocity $\dot{x}_i$ and $i=1,2,...,N$ ) includes all the parts coupled to system 1 by friction and receiving the dissipated mechanical energy. The energy transfer from system 1 to system 2 occurs only through a friction force. The total Hamiltonian reads $H=K_1+V_1+K_2+V_2+H_{int}$ where $K_1=\frac{1}{2}m\dot{x}^2$ is the kinetic energy and $V_1$ the potential energy of the system 1, $K_2=\frac{1}{2}\sum_{i}^{N}m_i\dot{x}_i(t)^2$ the kinetic energy and $V_2=\sum_{i}^{N}v(x_i)$ the potential energy of system 2 with $v(x_i)$ the potential energy of the particle $i$, and $H_{int}$ the interaction energy between the system 1 and 2. Since we are only interested in the paths or the equation of motion of the system 1, it is reasonable to suppose that system 2 does not move as a whole. This implies that $K_2$ and $V_2$ are only internal energies of the environment. $H_{int}$ is responsible for the friction law and determined by the coupling mechanism on the interface between the moving body and the environment. We can suppose that, for a limited time period of the motion under consideration, this interface (body's shape and size, body-environment distance, nature of the closest parts of the environment to the interface, etc.) and the friction law do not change significantly, hence the interaction mechanism should not change with the virtual variation of paths. So $H_{int}$ can be neglected in the following calculus, i.e., $H=K_1+K_2+V_1+V_2$ or $H=H_1+H_2$ where $H_1=K_1+V_1$ is the total energy of the system 1 and $H_2=K_2+V_2$ is the total energy of the system 2.

It should be stressed that the impact of the thermal fluctuation in system 2 on system 1 should be neglected. The motion of the system 1 remains classical mechanical and deterministic. This is reasonable for a mechanical body which is much larger than the constituents of system 2 and has much larger energy variation during the motion than the energy fluctuation of the thermal motion in system 2.

\section{Variational formulation of LAP for damped motion}
The Lagrangian $L$ of the whole conservative system can be defined by using the Legendre transformation $L=p\dot{x}+\sum_{i}^{N}p_i(t)\dot{x}_i(t)-H=K_1+2K_2-V_1-H_2$ where $p$ is the momentum of system 1 and $p_i$ the momentum of the particle $i$ of the system 2. The corresponding action is given by

\begin{equation}                                            \label{e5a}
A=\int_0^{t_b}Ldt=\int_0^{t_b}(\frac{1}{2}m\dot{x}^2+2K_2-V_1-H_2)dt.
\end{equation}
$H_2$ can be expressed as a function of $x_i(t)$ and the velocity $\dot{x}_i(t)$ of the constituents particles of the system 2. Its general expression reads
\begin{equation}                                            \label{e5aa}
H_2=\sum_{i}^{N}\frac{1}{2}m\dot{x}_i^2(t)+V_2[x_1(t),x_2(t)...x_N(t)]
\end{equation}
With this expression of $H_2$, the Lagrangian only depends on its variables at the time moment $t$, i.e.,
\begin{equation}                                            \label{e5aaa}
L=K_1(\dot{x}(t))+2K_2(\dot{x}_i(t))-V_1(x(t))-H_2(x_i(t),\dot{x}_i(t))\;\;\;i=1,2,...,N
\end{equation}
With this Lagrangian, the question of the nonlocality in time (see the remarks below) of the variational calculus, raised in the first formulation of the LAP for dissipative systems and discussed in detail in \cite{Wang2}, does not arise. The variation calculus can be made in the usual way as with the action of Eq.(\ref{e1}). A tiny path variation $\delta x(t)$ yields the following variation of the action:
\begin{equation}    \label{e5a1}
\begin{aligned}
\delta A&=\int_0^{t_b}[\frac{\partial L}{\partial x}\delta x+\frac{\partial L}{\partial\dot{x}}\delta\dot{x}]dt\\
&=\int_0^{t_b}[\frac{\partial L}{\partial x}-\frac{d}{dt}\frac{\partial L}{\partial\dot{x}}]\delta x dt.
\end{aligned}
\end{equation}
where we have made a time integration by parts of $\delta\dot{x}$ with the conditions $\delta x(a)=\delta x(b)=0$. The LAP requires $\delta A=0$, which, due to the arbitrary nature of $\delta x(t)$, leads to

\begin{equation} 	\label{e5a2}
\frac{\partial L}{\partial x}-\frac{d}{dt}\frac{\partial L}{\partial\dot{x}}=0.
\end{equation}
This is the differential equation for the damped motion of the system 1.

In order to use Eq.(\ref{e5a2}), the expression of $H_2$ in Eq.(\ref{e5aa}) must be changed into an explicit function of $x$ and $\dot{x}$. For this purpose, we consider the fact that $H_2$ changes in time only due to the energy dissipation of the system 1 and this dissipation is a function of $x$ and $\dot{x}$. We can write $H_2(t)=H_2(t_a)+E_d(x(t))$ where $H_2(t_a)$ is the energy of the system 2 at the initial time $t_a$, hence a constant for the motion hereafter, and $E_d(x(t))$ the part of the energy of system 1 dissipated by the friction force $f_d(x,\dot{x})$ from the initial moment $t_a$ to a moment $t$ ($0\leq t \leq t_b$):

\begin{equation}                                            \label{e3}
E_d(x(t))=-\int_{x_a}^{x(t)}f_d(x(\tau),\dot{x}(\tau))dx(\tau)
\end{equation}
where $\tau$ is any time moment between $t_a=0$ and $t$, and $dx(\tau)$ a small displacement along the path at the time $\tau$. The Lagrangian can be written as $L(x(t),\dot{x}(t))=K_1(\dot{x}(t))+2K_2(\dot{x}_i(t))-V_1(x(t))-E_d(x(t))$ where the constant $H_2(t_a)$ has been dropped. Introducing this Lagrangian into Eq.(\ref{e5a2}) and considering the fact that $K_1$ and $K_2$ are not explicit function of $x(t)$ and that $K_2$, $V_1$ and $E_d$ are not explicit function of $\dot{x}(t)$, one gets

\begin{equation} 	\label{e5a3}
\frac{d}{dt}\frac{\partial K_1}{\partial\dot{x}}+\frac{\partial V_1}{\partial x}+\frac{\partial E_d}{\partial x}=0.
\end{equation}
Considering $f_d(x(t),\dot{x}(t))=-\frac{\partial E_d(x(t))}{\partial x(t)}=-\frac{\partial}{\partial x(t)}\int_0^{x(t)}f(x(\tau),\dot{x}(\tau))dx(\tau)$ \cite{Calculus}, we straightforwardly obtain

\begin{equation} 	\label{e5a4}
m\ddot{x}=-\frac{\partial V_1}{\partial x}+f_d,
\end{equation}
the Newtonian equation for the damped motion with the friction $f_d$.

Here we remark that, at the stage of the variation calculus in Eq.(\ref{e5a1}), we have preferred not to replace $H_2$ with the functional $E_d(x(t))$ given by Eq.(\ref{e3}). The reason is that the time integral in Eq.(\ref{e3}) may be misleading and give a false impression that the energy $H_2$ and, consequently, the Lagrangian $L$ are nonlocal in space and time \cite{Wang2}, while $H_2$, in the expression Eq.(\ref{e5aa}) with its proper variables $x_i(t)$ and $\dot{x}_i(t)$, is completely local in time. This deceptive non locality of $H_2$ comes from the calculation of $E_d(x(t))$ by a cumulation of the work of the dissipative force $f_d$ over the configuration space as given in Eq.(\ref{e3}). As a matter of fact, not only $E_d$, any function of $x(t)$ or $\dot x(t)$ can surely be expressed as the time integral of another function of $x(t)$ or $\dot x(t)$. For instance, the position $x(t)=\int_ 0^t\dot{x}(\tau)d\tau$, the velocity $\dot x(t)=\int_ 0^t\ddot{x}(\tau)d\tau$, the kinetic energy $K_1(\dot{x}(t))=\int_0^t m\dot{x}(\tau)\ddot{x}(\tau)d\tau$, and the potential energy $V_1(x(t))=-\int_{x_0}^{x(t)}f_c(x(\tau))dx(\tau)=-\int_{0}^{t}f_c(x(\tau))\dot x(\tau)d\tau$ for a conservative force $f_c$. However, these expressions do not mean that, when a trajectory gets a certain variation, it is necessary, for the variation of those quantities, to consider $\delta x(\tau)$ prior to the moment $t$ in addition to $\delta x(t)$ at the moment $t$.

Indeed, with the non locality of $L$ or $H_2$ expressed in the integral Eq.(\ref{e3}), it seems reasonable to take into account the variation $\delta x(\tau)$ of the trajectory in addition to the variation $\delta x(t)$ \cite{Wang2}. But this operation leads to an incorrect equation of the motion as shown in the appendix of \cite{Wang2}. The problem is of course either in the Lagrangian or in the variational calculus. We are prone to say that the calculus including $\delta x(\tau)$ is not appropriate from physical point of view. The argument is the following.

On the one hand, the consideration of $\delta x(\tau)$ for the study of the motion at the moment $t$ is equivalent to considering the influence on the motion from an energy which has been dissipated before the moment $t$ and already dispersed in the environment. This is of course impossible, at least in classicla mechanics, if the thermal effect of the environment on the moving body is neglected as stated in the introduction.

On the other hand, a flaw of this variational technique including $\delta x(\tau)$ can be demonstrated by contradiction as follows. Write the kinetic energy $K_1$ as a cumulation of the work done by the inertia force $f_i=-m\ddot{x}$, one gets $K_1=-\int_{0}^{t}f_i(x(\tau))\dot x(\tau)d\tau=\int_0^t m\dot{x}(\tau)\ddot{x}(\tau)d\tau$ which is similar to Eq.(\ref{e3}). Put this expression into the variational calculus of Eq.(\ref{e1}) with the Lagrangian $L=K_1-V_1$ (without dissipation) and apply the technique of variation involving $\delta x(\tau)$ proposed in \cite{Wang2}, one will get
\begin{equation} 	\label{e10}
f_c=m\ddot{x}-m(t_b-t)\dddot{x}
\end{equation}
which has one more term $m(t_b-t)\dddot{x}$ (non zero in general) than the correct equation of motion. One can also replace the potential energy in the Lagrangian by the integral $V_1(x(t))=-\int_{0}^{t}f_c(x(\tau),\dot{x}(\tau))\dot x(\tau)d\tau$, the equation of motion will be still more whimsical if $f_c$ explicitly depends on velocity as in the case of a charged body moving in a magnetic field.

From the above analysis, it is clear that the variation calculus of LAP should be limited to the moment $t$ by considering only $\delta x(t)$ and $\delta\dot x(t)$ when the Lagrangian, dissipative or not, includes non local expression of energy involving other moment than $t$. From physical point of view, this rule is equivalent to saying that the state of motion of a body at time $t$ is only affected by its own energy changes at the same moment $t$. Of course, the best way to avoid confusion and polemics \cite{Wang2} is to write the Lagrangian, if possible, in the form of a local function in space and time, as we have done in Eq.(\ref{e5aaa}) when writing $H_2$, instead of $E_d$, in the Lagrangian and making the variation in Eq.(\ref{e5a1}).

\section{Derivation from virtual work principle}
The above LAP can also be derived from other fundamental principles as has been done in analytical mechanics by using virtual work principle of d'Alembert\cite{Arnold,Alembert}. This latter principle is valid in the presence of friction force. For 1-dimensional moving body, it reads:
\begin{equation}                                            \label{e4}
\delta W=(f_c+f_d-m\ddot{x})\delta x=0
\end{equation}
where $f_c=-\frac{\partial V_1}{\partial x}$ is the conservative force. Using $\ddot{x}\delta x=\frac{d}{dt}(\dot{x}\delta x)-\dot{x}\delta\dot{x}=\frac{d}{dt}(\dot{x}\delta x)-\frac{\partial K_1}{\partial \dot{x}}\delta\dot{x}$, then integrating Eq.(\ref{e4}) over time from $t_a=0$ to $t_b$, we get
\begin{equation}                                            \label{e6}
\int_0^{t_b}(\frac{\partial K_1}{\partial \dot{x}}\delta\dot{x}-\frac{\partial (V_1+E_d)}{\partial x}\delta x)dt=\int_0^{t_b}(\frac{\partial L}{\partial \dot{x}}\delta\dot{x}+\frac{\partial L}{\partial x}\delta x)dt=\int_0^{t_b}\delta Ldt=\delta A=0
\end{equation}
where we used $\delta x(t_a)=\delta x(t_b)=0$, $L=K_1-V_1-E_d$ and $A=\int_0^{t_b}Ldt$.

\section{Concluding remarks}
In summary, we have proposed a possible formulation of the least action principle for damped motion with an universal, energy connected and unique Lagrangian. We have used the model of a whole conservative system composed of the damped moving body and of its environment receiving the dissipated energy. It is also shown that this formulation can be derived from the virtual work principle. We hope that these results are helpful for further study of the relations between the fundamental principles of Lagrangian/Hamiltonian mechanics and the variational principles relative to energy dissipation \cite{Liu,Onsager,Eyink,Prados,Kodama}.

This work is presented in term of friction as the coupling between the moving body and its environment. But the formulation is not limited to friction. This is an advantage of the substitution of $H_2$ by $E_d$ which is written as a function of the damped motion without involving the details of the motions of the tiny particles in the environment. Hence the coupling can be any mechanism depending on the motion and dissipating energy from it. For instance, the emission of electromagnetic wave or light into the void (environment) from an accelerated charged body, or emission of sound by a vibrating body.

Finally we would like to mention that, although the term ``least action'' is used here for historical reason, the stationarity $\delta A=0$ is not necessarily a minimum. The nature of the stationarity (minimum, maximum or inflection) of $A$ has been addressed in our recent work\cite{Wang2} by numerical simulation of damped motion and of variational analysis, in which the action $A_{op}$ along the optimal path is calculated and compared to the actions of many paths around the optimal one. One of the conclusions is that the stationarity of $A_{op}$ undergoes a transition from minimum to maximum when drag constant $\zeta$ and the dissipated energy increase. Further investigation is necessary to confirm this evolution of the stationarity of the action.

\section*{Acknowledgments}
This work was supported by the Region des Pays de la Loire of France under the Grant No.2009-9397.

\end{document}